\documentclass[reprint,superscriptaddress,amsmath,amssymb,aps,prd]{revtex4-2}

\usepackage{bm}
\usepackage{amsthm}
\usepackage{amssymb}
\usepackage{amsfonts}
\usepackage{eucal}
\usepackage{mathrsfs}
\usepackage{braket}
\usepackage{xcolor}
\usepackage{graphicx}
\usepackage{booktabs}
\usepackage[T1]{fontenc}
\usepackage[utf8]{inputenc}
\usepackage{textcomp}
\usepackage[hidelinks]{hyperref}
\usepackage{newunicodechar}
\usepackage{subcaption}
\usepackage{orcidlink}
\newunicodechar{̈}{\"{}}

\newcommand{\tr}{\text{Tr}}
\def\wo{{\omega_0}}
\def\wk{{\omega_k}}
\def\wj{{\omega_j}}
\def\bk{{\bf k}}
\def\wku{{\omega_k^{(1)}}}
\def\wkd{{\omega_k^{(2)}}}
\def\wkm{{\omega_k^{(\mu )}}}

\def\wju{{\omega_j^{(1)}}}
\def\wjd{{\omega_j^{(2)}}}
\def\wjm{{\omega_j^{(\mu )}}}

\def\wlu{{\omega_\ell^{(1)}}}

\def\wlm{{\omega_\ell^{(\mu )}}}
\def\wln{{\omega_\ell^{(\nu )}}}

\def\wmd{{\omega_m^{(2)}}}

\def\wmn{{\omega_m^{(\nu )}}}

\def\ckju{{C_{kj}^{(1)}}}
\def\ckjd{{C_{kj}^{(2)}}}
\def\ckjm{{C_{kj}^{(\mu )}}}
\def\cljm{{C_{\ell j}^{(\mu )}}}

\def\clmn{{C_{\ell m}^{(\nu )}}}

\def\clku{{C_{\ell k}^{(1)}}}
\def\ckku{{C_{k k}^{(1)}}}
\def\cmjd{{C_{mj}^{(2)}}}
\def\cjjd{{C_{jj}^{(2)}}}

\def\aku{{a_k^{(1)}}}
\def\aju{{a_j^{(1)}}}
\def\akd{{a_k^{(2)}}}
\def\ajd{{a_j^{(2)}}}
\def\akud{{a_k^{(1)\dagger}}}
\def\ajud{{a_j^{(1)\dagger}}}
\def\akdd{{a_k^{(2)\dagger}}}
\def\ajdd{{a_j^{(2)\dagger}}}

\def\akmd{{a_k^{(\mu )\dagger}}}
\def\ajmd{{a_j^{(\mu )\dagger}}}
\def\almd{{a_\ell^{(\mu )\dagger}}}

\def\aln{{a_\ell^{(\nu )}}}
\def\amn{{a_m^{(\nu )}}}
\def\alnd{{a_\ell^{(\nu )\dagger}}}
\def\amnd{{a_m^{(\nu )\dagger}}}

\def\vac{\ket{0; \{ 0 \}_1, \{ 0 \}_2}}

\begin{document}

\title{Entanglement generation between field modes mediated by a fluctuating conducting wall}

\author{Luca Giovanni Cammarata \orcidlink{0009-0009-4126-2983}}
\email{lucagiovanni.cammarata@unipa.it}
\noaffiliation
\affiliation{Dipartimento di Fisica e Chimica - Emilio Segr\`e, Universit\`a degli Studi di Palermo, Via Archirafi 36, I-90123 Palermo, Italy}
\author{Tommaso Fazio \orcidlink{0000-0003-3165-1095}}
\email{tommaso.fazio@unipa.it}
\noaffiliation
\affiliation{Dipartimento di Fisica e Chimica - Emilio Segr\`e, Universit\`a degli Studi di Palermo, Via Archirafi 36, I-90123 Palermo, Italy}
\author{Roberto Passante \orcidlink{0000-0003-4884-0028}}
\email{roberto.passante@unipa.it}
\noaffiliation
\affiliation{Dipartimento di Fisica e Chimica - Emilio Segr\`e, Universit\`a degli Studi di Palermo, Via Archirafi 36, I-90123 Palermo, Italy}
\author{Lucia Rizzuto \orcidlink{0000-0001-8556-5542}}
\email{lucia.rizzuto@unipa.it}
\noaffiliation
\affiliation{Dipartimento di Fisica e Chimica - Emilio Segr\`e, Universit\`a degli Studi di Palermo, Via Archirafi 36, I-90123 Palermo, Italy}

\date{\today}

\begin{abstract}
\noindent We consider a movable conducting plate of finite mass, between two fixed ones, whose mechanical degrees of freedom are treated quantum-mechanically and bound to its equilibrium position by a harmonic potential. The movable wall is thus subjected to quantum fluctuations of its position. This creates a system of two subcavities separated by the movable fluctuating plate, and two massless one-dimensional scalar fields, one in each subcavity. This system is described by an appropriate generalization of the Law Hamiltonian. The presence of the movable wall yields an effective plate-fields interaction, as well as an effective interaction between the field modes. We obtain, at the second order in perturbation theory, the ground state of the interacting system and the reduced density operator of the fields in each subcavity by tracing out the wall's degrees of freedom. We calculate the entanglement between two field modes, one in each cavity, by evaluating analytically the negativity; we then also evaluate numerically the total multimode negativity. Our results show that in both cases the fields in the two subcavities are entangled, in contrast to the case in which the wall is fixed in space. We discuss the amount of the field entanglement present as a function of relevant physical parameters of the system such as the mass and oscillation frequency of the movable wall, its distance from the fixed walls and the frequencies of the field modes considered.
\end{abstract}

\maketitle

\section{Introduction}

The presence of a reflecting or dielectric wall in space gives boundary conditions on quantum fields, that in general can modify the allowed field modes and the resulting mode spectrum. In the common case of a perfectly reflecting infinite plane boundary with a fixed position, the boundary separates the space into two completely independent half-spaces. The situation is different when the boundary is allowed to move. In such a case, two conceptually different cases are possible: the boundary has a prescribed motion in space, and this motion is treated classically, or the boundary's motion is described quantum mechanically and its degrees of freedom are associated to quantum operators included in the total Hamiltonian of the system considered.

These situations, namely a prescribed classical motion (for instance oscillatory) or a quantum mechanical motion of the boundary,
including possible dissipative and backreaction effects, are typical of quantum optomechanics \cite{Aspelmeyer-Kippemberg14} and of the dynamical Casimir effect \cite{Dodonov20,MartinCaro-GarciaMoreno24}; several new physical phenomena occur in such a case. Optomechanical systems have received strong interest in the literature also as systems for exploring possible tests of fundamental physics and gravity modifications to quantum mechanics \cite{Chen13}. The generation of entanglement in an optomechanical system has also been discussed in recent literature \cite{andreata2005dynamics,velasco2022photon,del2020entanglement}, as well as modifications of radiative processes of quantum emitters embedded in dynamical (time-dependent) media \cite{Calajo-Rizzuto17,Bello-AsgarnezhadZorgabad26}.

The case of a movable boundary whose motion is described quantum mechanically, and thus subjected to quantum position fluctuations, yields additional relevant phenomena \cite{Aspelmeyer-Kippemberg14} that can be described through appropriate effective Hamiltonians that include the interaction between the movable wall and quantum fields; indeed, an effective field-wall interaction energy is found as well as an effective interaction among field modes mediated by the movable wall (Law Hamiltonian) \cite{Law95,Law94}.
This latter case has been recently considered in the literature, for example a one-dimensional cavity consisting of a fixed and a movable wall bound to its equilibrium position by a harmonic potential: field energy density inside the cavity in the ground state, corrections to the Casimir interaction between the two walls \cite{Butera-Passante13,Armata-Passante15}, as well as dynamical self-dressing of the movable wall, starting from a nonequilibrium configuration \cite{Armata-Kim17}, have been investigated.
A similar effective Hamiltonian has also been used for studying the quantum radiation emitted by shaking an atom in the vacuum \cite{Lo-Law18}. Quantum friction effects on the mechanical motion of the mirror due to dynamical Casimir emission \cite{Butera-Carusotto19} and higher-order effects in the position fluctuations of the movable mirror on the radiation pressure \cite{Butera25} have also been considered.
Quantum entanglement between mechanical oscillators or between populated single-mode electromagnetic fields mediated by a mechanical oscillator has been investigated in the literature (for a review, see \cite{Tang-Cai22}), as well as
multiphoton entanglement in a single-mode case \cite{10.21468/SciPostPhys.18.2.067} or optomechanical heat transfer \cite{Zhai:26}. Recently, ground-state correlations between field observables (specifically, between field squared operators) at the two sides of a movable conducting mirror have been evaluated, using an appropriate generalization of the Law Hamiltonian \cite{Montalbano-Armata23,Armata-Butera23}. These results highlight that even a perfectly conducting mirror of a finite mass, whenever moving in space (a fluctuating position, for instance), allows quantum fields at its two sides to influence each other. This is also conceptually similar to recent proposals of fuzzy event horizons of a black hole in quantum gravity models \cite{Arias-Krein12,Giugno-Giusti18,Mertens-Turiaci23}. Furthermore, fluctuating boundaries have also been shown to be able to smear out divergences of field energy densities occurring at the position of fixed perfect boundaries \cite{Ford-Svaiter98,Butera-Passante13}.

In this paper we consider a general setup with a movable perfect mirror placed in an arbitrary position between two fixed perfect mirrors, as shown in Fig. \ref{fig:entanglement_mediated_by_flucwall}, bound to its equilibrium position by a harmonic potential, and two massless one-dimensional scalar fields defined in the two subcavities at both sides of the movable mirror. After generalizing the Law Hamiltonian to this setup, we obtain by perturbation theory the interacting ground state of the system at the second order in the field-wall effective interaction; this state contains terms with up to four virtual quanta in the fields and up to two excitations in the wall's motion. We then obtain the field reduced density operator and show the presence of entanglement between field modes at the two opposite sides of the movable mirror by evaluating the negativity as a measure of the entanglement.
An upper cutoff frequency is introduced to cure the ultraviolet divergences and to take into account that a real conducting boundary becomes transparent at frequencies larger than the plasma frequency of its material.
Both analytical and numerical evaluation and physical discussion of the negativity are performed.
Specifically, our results indicate the presence of intercavity field entanglement even in the absence of injected photons, hence
arising from the ground-state dressing of the system induced by the interaction of the vacuum field with the fluctuating wall.

This paper is organized as follows. In Sec. \ref{sec:model} we introduce our model, calculate the interacting ground state at the second order in the effective wall-fields interaction, and the field reduced density operator. In Sec. \ref{sec:entanglement} we evaluate the negativity between two arbitrary field modes, one in each subcavity, and discuss the entanglement between them and its dependence from the relevant physical parameters of the system. In Sec. \ref{sec:entanglement2} we evaluate numerically the multimode negativity between the field in the two subcavities. Finally, Sec. \ref{sec:conclusions} is devoted to our conclusive remarks.

\section{The model and the interacting ground state}
\label{sec:model}

We consider a direct generalization of a previously introduced one-dimensional model consisting of two quantum one-dimensional (1D) massless scalar fields defined at the two sides of a perfectly conducting movable wall. The movable wall, whose degrees of freedom are treated quantum-mechanically and included in the Hamiltonian, is subjected to quantum fluctuations of its position and it is bound around its equilibrium position by a harmonic potential \cite{Montalbano-Armata23,Armata-Butera23}. The Hamiltonian used is a direct two-cavity generalization of the Law model and Hamiltonian \cite{Law95,Law94}, that has been widely employed in the literature.
A related optomechanical two-cavity Hamiltonian was derived in \cite{russo2023optomechanical} starting from the quantization of the classical problem; the subsequent analysis was then specialized to the single-mode case in each subcavity to study two-photon hopping.
In the configuration investigated in \cite{Montalbano-Armata23,Armata-Butera23}, the average position of the fluctuating mirror coincides with the midpoint of a cavity delimited by two fixed perfectly conducting walls.
The physically relevant limit of the two external walls going to infinity, yielding a continuum of field modes, was then considered. In that context, correlations between field modes, mediated by the mirror position fluctuations, were analyzed together with their possible observability.

In this paper we consider a more general situation and address a different physical problem, specifically the generation of entanglement between field modes in the two half-spaces at the opposite sides of the movable wall.

\begin{figure}[h!]
    \centering
    \includegraphics[width=0.5\textwidth]{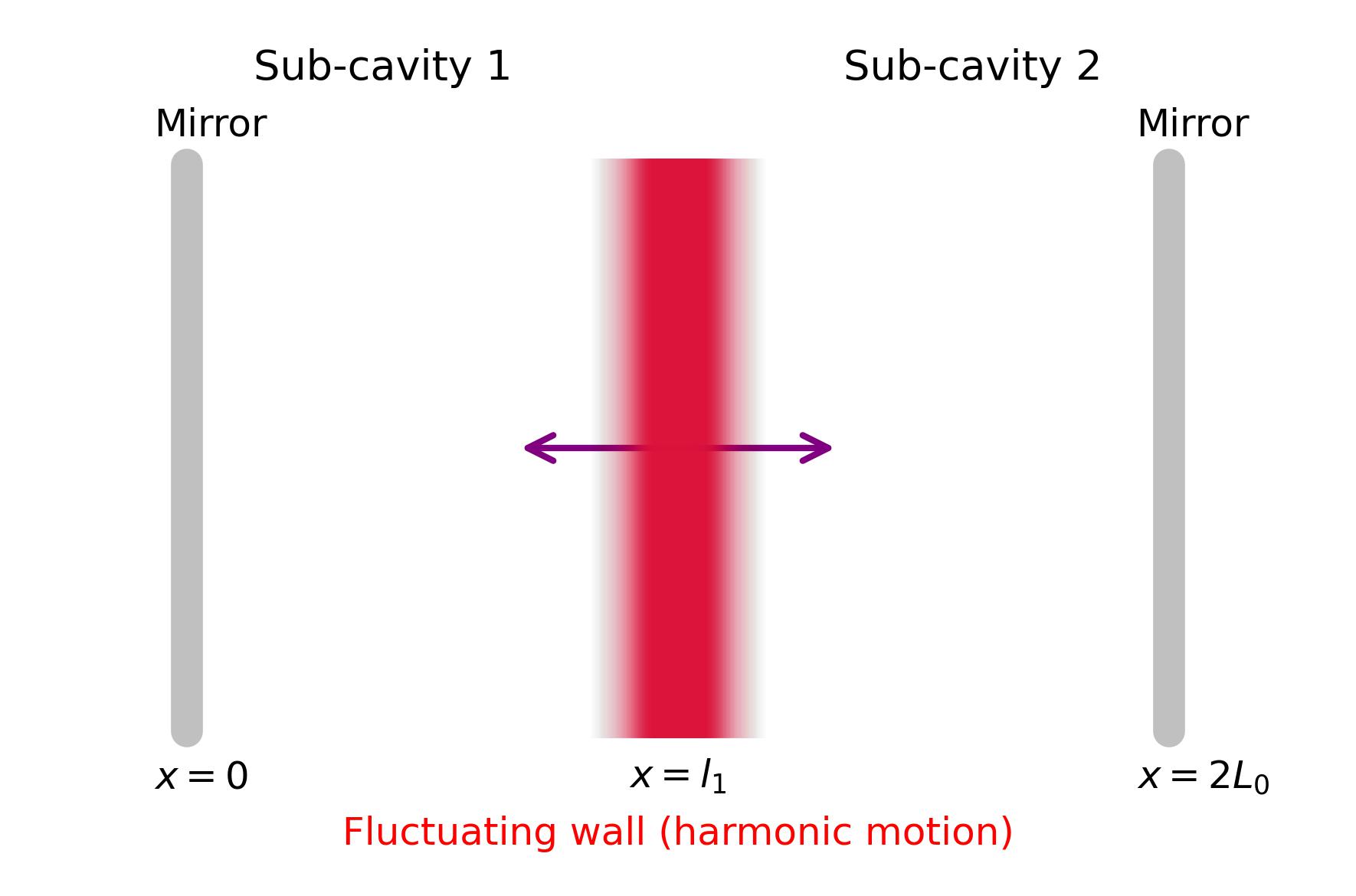}
    \caption{Pictorial representation of a cavity divided into two subcavities by a fluctuating conducting wall with equilibrium position between the two external walls.}
    \label{fig:entanglement_mediated_by_flucwall}
\end{figure}

Our system, illustrated in Fig. \ref{fig:entanglement_mediated_by_flucwall}, consists of two fixed conducting walls at $x=0$ and $x=2L_0$ and a movable conducting wall whose equilibrium position $l_1 \in (0,2L_0)$ lies between them.
The distances between the movable wall and the two fixed ones are therefore $l_1$ and $l_2 = 2L_0-l_1$, respectively. Our setup thus consists of two 1D cavities corresponding to the regions $(0,l_1)$ and $(l_1,2L_0)$, each containing a massless scalar field.
The (middle) movable mirror has finite mass $M$ and is bound to its equilibrium position $l_1$ by a harmonic potential of frequency $\omega_0$. Since its mechanical degrees of freedom are treated quantum mechanically and included in the Hamiltonian of the system, it exhibits quantum fluctuations of its position around the equilibrium position.

As it is known, the presence of the movable middle mirror generates, for both fields, an effective field-mirror interaction and an effective interaction between field modes \cite{Law95,Montalbano-Armata23}. This Hamiltonian was originally introduced in \cite{Law94,Law95} for a 1D system consisting of one fixed and one movable wall and a massless scalar field, and later extended to the case of two cavities composed by a movable wall at the midpoint between two fixed walls, and two scalar fields in each cavity \cite{Montalbano-Armata23}.
The Law effective Hamiltonian is obtained at the leading order in a series expansion of the wall's displacement from its equilibrium position with respect to the cavity size $\sim L_0$.
Here, we first extend this model to the configuration where the movable-wall equilibrium position can be any point between the two fixed walls, so that the two 1D cavities can have different lengths.

The effective Hamiltonian for our system is
\begin{equation}\label{Hamiltonian}
H=H_0 + H_I^1 + H_I^2 ,
\end{equation}
where
\begin{equation}\label{Unperturbed Hamiltoniam}
H_0 = \hbar \wo \, b^\dagger b +\hbar \sum_k \wku \akud \aku +\hbar \sum_k \wkd \akdd \akd
\end{equation}
is the unperturbed term. Here, $\wo = 2\pi \nu_0$ is the wall's oscillation angular frequency, $\wku = \pi c k/l_1$ and $\wkd = \pi c k/(2L_0-l_1)$ ($k=1,2,\ldots$) are the angular frequencies of the field modes in the first and in the second cavity, respectively; $\aku , \akud$ and $\akd , \akdd$ are the field annihilation and creation operators pertinent to the first and the second cavity, respectively, defined for field modes relative to the equilibrium position of the movable wall \cite{Law95}; they satisfy the usual bosonic commutation rules (also, operators relative to different cavities commute with each other, being independent degrees of freedom). $H_I^1$ and $H_I^2$ are the two effective coupling Hamiltonians between the movable wall and the scalar fields defined in cavities $1$ and $2$, respectively. They have the following form
\begin{equation}\label{Interaction Hamiltonian}
\begin{split}
   H_I^1 =& -(b+b^\dagger ) \sum_{kj} \ckju \text{N} \left[ \left(\aju +\ajud \right)\left(\aku +\akud \right) \right] , \\
   H_I^2 =& -(b+b^\dagger ) \sum_{kj} \ckjd \text{N} \left[ \left(\ajd +\ajdd \right)\left(\akd +\akdd \right) \right] ,
\end{split}
\end{equation}
N being the normal ordering operator, $k,j=1,2,\hdots$, and the coupling constants are
\begin{equation}\label{Coupling constant}
\begin{split}
\ckju =&  (-1)^{j+k} \left( \frac \hbar 2 \right)^{\frac 32} \frac 1{l_1} \sqrt{\frac {\wju \wku}{M \, \wo}}  , \\
\ckjd =& -(-1)^{j+k} \left( \frac \hbar 2 \right)^{\frac 32} \frac 1{2L_0 - l_1} \sqrt{\frac {\wjd \wkd}{M \, \wo}}  ,
\end{split}
\end{equation}
where $M$ is the mass of the movable mirror. We take this Hamiltonian as a general effective Hamiltonian model, describing an interaction linear in the mirror coordinate and quadratic in the field variables.
In principle, this Hamiltonian is derived under the assumption that both the movable and the two fixed mirrors are ideal conductors; thus, the field operators are required to satisfy boundary conditions that force them to vanish at the mirrors' locations. This assumption is of course good for field modes with frequencies below the plasma frequency of the metal of which the mirrors are actually made, since, above this threshold, the mirrors become transparent to the radiation. In particular, the movable wall ceases to separate the two half-spaces for modes with frequency above the plasma frequency \cite{Montalbano-Armata23}, that thus do not contribute to the processes under consideration here.  Therefore, we include in our Hamiltonian only a finite number of modes, $N_{\text{mod}}$, using an appropriate sharp frequency cutoff of the order of a typical metal plasma frequency $\omega_P$ (for the sake of simplicity, when there is not possibility of confusion we use just the word frequency in place of angular frequency); this number is determined by the relation
\begin{equation}\label{number modes}
N_{\text{mod}} \sim \frac{ \omega_P L_0}{\pi c} .
\end{equation}
As we shall see later on, however, our main result in Sec. \ref{sec:entanglement} does not depend on the cutoff frequency, and is finite for very large $N_{\text{mod}}$.
We write the eigenstates of $H_0$ as Fock states in the form $\ket{n_\text{wall}; \{ n_1 \}_1, \{ n_2 \}_2} = \ket{n_\text{wall}}\ket{\{ n_1 \}_1, \{ n_2 \}_2}$, where the first element refers to the wall's excitations, while the second and the third ones refer, respectively, to excitations of field modes in the first and second subcavity.
A similar Hamiltonian, describing the interaction between a mechanical mode and two optical modes in a resonant approximation, has recently been used for a detection proposal of dark matter using superfluid helium in an optomechanical cavity \cite{PhysRevD.110.043005}.

We use the following standard definition and normalization for multiquanta states in the Fock space for each subcavity \cite{Schweber05}
\begin{equation}
\label{FockStates}
\begin{split}
\ket{\bk_1 \ldots \bk_n} =& \frac 1{\sqrt{n!}} a^\dagger (\bk_1) \ldots a^\dagger (\bk_n) \ket{0} , \\
\braket{\bk_1' \ldots \bk_n'| \bk_1 \ldots \bk_n} &= \frac 1{n!} \sum_P \delta (\bk_1-\bk_{\alpha_1}') \ldots \delta (\bk_n-\bk_{\alpha_n}'),
\end{split}
\end{equation}
where the sum is over all permutations $P$ of the set $\{ \alpha_1, \ldots, \alpha_n \}$, and $\ket{0}$ is the (normalized) vacuum state of the field.

The unperturbed ground state of the system is $\ket{g_0} = \vac$, with no excitations in the wall and in both scalar fields. The interacting ground state is obtained by perturbation theory up to the second order
\begin{equation}
    \ket{\tilde{g}} = \left( 1 - \Lambda^2 \right) \ket{g_0} + \ket{g_1} + \ket{g_2}  ,
\end{equation}
where $\ket{g_1}$ is the first-order correction in the effective wall-cavities interactions, $\ket{g_2}$ is the second-order one and $\Lambda^2$ is a normalization factor arising from the second-order correction.
The first-order correction is
\begin{equation}
\label{first-order correction}
\ket{g_1} = \sum_{\mu =1,2} \sum_{jk} \frac {\ckjm}{\hbar (\wo +\wkm +\wjm )}
\akmd \ajmd \ket{ 1; \{ 0 \}_1, \{ 0 \}_2} ,
\end{equation}
containing two virtual excitations in one of the two cavities and one mirror excitation. The second-order normalization factor is
\begin{equation}\label{Normalization factor}
  \Lambda^2= \sum_{\mu =1,2}\sum_{jk} \frac {(\ckjm )^2}{\hbar^2(\wo +\wjm +\wkm)^2} ,
\end{equation}
where an upper frequency cutoff in the sums over $j$ and $k$ is assumed.
The second-order correction, apart from the normalization factor (\ref{Normalization factor}), is given by
\begin{equation}
\label{second-order correction}
\ket{g_2} = \ket{g_2}_0 + \ket{g_2}_1 + \ket{g_2}_2 ,
\end{equation}
where $\ket{g_2}_0, \ket{g_2}_1, \ket{g_2}_2$ are states containing respectively zero, one and two pairs of total excitations in the two cavity fields. They are given by
\begin{widetext}
\begin{equation}\label{second-order states}
  \begin{split}
  \ket{g_2}_0 &=  \frac {\sqrt{2}}{\hbar^2 \wo} \sum_{\mu =1,2}\sum_{kj} \frac {(\ckjm )^2}{\wo + \wjm +\wkm} \ket{2; \{ 0 \}_1, \{ 0 \}_2} , \\
  \ket{g_2}_1 &= \frac {4}{\hbar^2} \sum_{\mu =1,2} \sum_{jk\ell} \frac {\ckjm \cljm}{\wo + \wjm +\wkm} \akmd \almd \left[ \frac {1}{\wkm +\wlm} \ket{ 0; \{ 0 \}_1, \{ 0 \}_2}
  + \frac {\sqrt{2}}{2\wo +\wkm +\wlm} \ket{ 2; \{ 0 \}_1, \{ 0 \}_2} \right] ,  \\
      \ket{g_2}_2 &=    \frac 1{\hbar^2} \sum_{\mu \nu =1,2} \sum_{jklm} \frac {\ckjm}{\wo + \wjm +\wkm}
      \left[ \frac {\clmn}{\wjm +\wkm +\wln +\wmn} \ket{0} + \frac {\sqrt{2}\clmn}{2\wo +\wjm +\wkm +\wln +\wmn} \ket{2} \right] \\
      &\ \ \ \times \alnd \amnd \akmd \ajmd \ket{\{ 0 \}_1, \{ 0 \}_2} .
  \end{split}
\end{equation}
\end{widetext}

The density operator for the complete system (pure state), up to the second order in the coupling constants, is
\begin{equation}\label{density operator}
    \begin{split}
        \rho &= \ket{\tilde{g}} \bra{\tilde{g}} = (1 - 2 \Lambda^2) \ket{g_0} \bra{g_0} + \ket{g_1} \bra{g_1} \\
       \quad \quad &+ \left( \ket{g_0} \bra{g_1} + \ket{g_0} \bra{g_2} + \text{H.c.} \right) .
    \end{split}
\end{equation}

The second-order reduced field density operator is obtained by tracing out the wall's degrees of freedom
\begin{widetext}
\begin{equation} \label{reduced density operator}
    \begin{split}
    \rho_F &= \tr_w \lbrace \rho \rbrace = \left( 1-2\Lambda^2 \right) \ket{\{ 0 \}_1, \{ 0 \}_2} \bra{\{ 0 \}_1, \{ 0 \}_2} \\
    &+ \frac {1}{\hbar^2} \left[ \sum_{\mu \nu =1,2} \sum_{jk\ell m}
    \frac {\ckjm \clmn}{(\wo +\wkm +\wjm )(\wo +\wln + \wmn )} \akmd \ajmd \ket{\{ 0 \}_1, \{ 0 \}_2} \bra{\{ 0 \}_1, \{ 0 \}_2} \aln \amn \right. \\
    &+ 4 \sum_{\mu =1,2} \sum_{jk\ell} \frac {\ckjm \cljm}{(\wo +\wkm +\wjm )(\wkm + \wlm )} \akmd \almd \ket{\{ 0 \}_1, \{ 0 \}_2} \bra{\{ 0 \}_1, \{ 0 \}_2}  \\
    &\ \left. +\sum_{\mu \nu =1,2} \sum_{jk\ell m} \frac {\ckjm \clmn}{(\wo +\wkm +\wjm )(\wjm +\wkm +\wln + \wmn )} \akmd \ajmd \alnd \amnd \ket{\{ 0 \}_1, \{ 0 \}_2} \bra{\{ 0 \}_1, \{ 0 \}_2}
    + \text{H.c.} \right] ,
    \end{split}
\end{equation}
\end{widetext}
where $\tr_w$ indicates the partial trace operation over the degrees of freedom of the wall. From Eq. (\ref{reduced density operator}), the purity can be obtained as
$\mathcal{P} (\rho_F) = \tr \lbrace \rho_{F}^2 \rbrace =1 - 4 \Lambda^2 < 1$, showing that, as expected, $\rho_{F}$ is a mixed state.
Also, we must have $1-4\Lambda^2 >0$, and this sets an upper limit to $\Lambda^2$, which depends also on the cutoff frequency; we will discuss this important point in the next section, showing that indeed, for typical experimental values of all parameters involved in our system, we find $\Lambda^2 \ll 1$ and thus our perturbative calculation is consistent.
Also, the von Neumann entropy cannot be used as a proper measure of the degree of entanglement between the fields in the two subcavities, since its interpretation as an entanglement quantifier is valid only for globally pure field states \cite{CohenTannoudji-Diu20}. We therefore adopt a different entanglement estimator, namely the negativity \cite{vidal2002computable}. To introduce it, let $\mathcal{H}_a$ and $\mathcal{H}_c$ be two Hilbert spaces, and $O$ an operator in the tensor-product space $\mathcal{H}_a \otimes \mathcal{H}_c$. If $\{ \varphi_j \}$ is an orthonormal basis in $\mathcal{H}_a$ and $\{ \psi_m \}$ an orthonormal basis in $\mathcal{H}_c$, we can decompose $O$ as
\begin{equation}
    \begin{split}
        O &= \sum_{jkmn} O_{jmkn} \ket{\varphi_j , \psi_m} \bra{\varphi_k, \psi_n} \\
        & = \sum_{jkmn} O_{jmkn} \ket{\varphi_j} \bra{\varphi_k} \otimes \ket{\psi_m} \bra{\psi_n} \, ,
    \end{split}
\end{equation}
where $O_{jmkn} = \langle \varphi_j , \psi_m |O| \varphi_k, \psi_n \rangle $. The partial transpose of $O$ with respect to the subspace "$c$" is defined as
\begin{equation}
    \begin{split}
        O^{t_c} &= \sum_{jkmn} O_{jmkn} \ket{\varphi_j} \bra{\varphi_k} \otimes (\ket{\psi_m} \bra{\psi_n})^\dagger \\
        & =\sum_{jkmn} O_{jmkn} \ket{\varphi_j} \bra{\varphi_k} \otimes \ket{\psi_n} \bra{\psi_m} \\
        & = \sum_{jkmn} O_{jnkm} \ket{\varphi_j} \bra{\varphi_k} \otimes \ket{\psi_m} \bra{\psi_n} \, .
    \end{split}
\end{equation}
The negativity of a state $\rho$ is then defined as \cite{vidal2002computable}
\begin{equation}
    \mathcal{N} (\rho) = \frac{ \tr  \lbrace \sqrt{(\rho^{t_c})^\dagger \, \rho^{t_c}} \,  \rbrace\, -1}{2} = \Big| \sum_{\lambda  <0} \lambda  \, \Big| \, ,
\end{equation}
where the real numbers $\lambda$ are the eigenvalues of $\rho^{t_c}$.

The reduced density operator derived above, together with the definition of the negativity, provides the starting point for our quantitative analysis of the entanglement generated between field modes in the two subcavities, that will be addressed in the next section.

\section{Quantification of the entanglement between a pair of modes in different cavities}
\label{sec:entanglement}

In order to quantify the entanglement between the scalar fields in the two subcavities through the negativity, we now evaluate the negativity associated with the reduced density operator describing the joint state of two field modes, one for each cavity, rather than the negativity of the complete field state $\rho_F$ (which cannot be diagonalized analytically). This allows us to obtain analytical results for the negativity.
The relevant reduced density operator is obtained by fixing two field modes, say $k$ in the first cavity and $j$ in the second one, and then tracing out all the other field modes.
The modes $j$ and $k$ are chosen such that their frequencies are far less than the relevant cutoff frequency of the theory.
Thus, we define the reduced density operator
\begin{equation}
    \begin{split}
        \rho_{kj} &=  \tr'_{kj}  \lbrace \rho_F  \rbrace \\
        &= \sum_{\{n_{l}^{(1)} \} : l \ne k } \sum_{ \{n_{m}^{(2)} \} : m \ne j } \langle n_{l}^{(1)} , n_{m}^{(2)} |\rho_F | n_{l}^{(1)} , n_{m}^{(2)} \rangle \, ,
    \end{split}
\end{equation}
where $\tr'_{kj}$ denotes a partial trace over all field modes but the mode $k$ of subcavity $1$ and the mode $j$ of subcavity $2$.

After some lengthy but straightforward calculations, we obtain

\begin{widetext}
    \begin{equation}\label{stato campi due modi}
    \begin{split}
        \rho_{kj} &= \Bigg[ 1-\frac{2}{\hbar^2} \Bigg(
        \sum_{\ell}\frac{\left(C^{(1)}_{\ell k}\right)^2 (2-\delta_{\ell k})}{\left(\omega_0+\omega^{(1)}_\ell+\omega^{(1)}_k\right)^2}+
        \sum_{m}\frac{\left(C^{(2)}_{mj}\right)^2 (2-\delta_{mj})}{\left(\omega_0+\omega^{(2)}_m+\omega^{(2)}_j\right)^2} \Bigg)\Bigg]
         \ket{0_k , 0_j} \bra{0_k , 0_j}
        \\
        &\ \ +\frac {1}{\hbar^2} \Bigg\{ 4\sum_{l \ne k} \frac{ (C_{lk}^{(1)} )^2}{( \omega_0 + \omega^{(1)}_k + \omega^{(1)}_l )^2} \ket{ 1_k , 0_j} \bra{1_k , 0_j}
        + 4\sum_{m \ne j} \frac{ ( C_{mj}^{(2)} )^2 }{(\omega_0 + \omega^{(2)}_j + \omega^{(2)}_m )^2} \ket{ 0_k , 1_j} \bra{0_k , 1_j} \\
         &\ \ \ +  2\sqrt{2} \Big[ \sum_\ell \frac{(\clku )^2}{\wku (\wo + \wlu +\wku)} \big( \ket{2_k , 0_j} \bra{0_k , 0_j} + \ket{0_k , 0_j } \bra{2_k, 0_j} \big) \\
         &\ \ \ + \sum_{m} \frac {\left( \cmjd \right)^2}{\wjd (\wo + \wmd +\wjd)} \big( \ket{0_k , 2_j} \bra{0_k , 0_j} + \ket{0_k , 0_j} \bra{0_k , 2_j} \,  \big) \Big] + \frac {2(\ckku )^2}{(\wo + 2\wku )^2} \ket{2_k , 0_j} \bra{2_k, 0_j} \\
         &\ \ \ + \frac {2(\cjjd )^2}{(\wo + 2\wjd )^2} \ket{0_k , 2_j} \bra{0_k, 2_j}
         + \frac {2\ckku \cjjd}{(\wo + 2\wku )(\wo + 2\wjd )} \big( \ket{2_k , 0_j} \bra{0_k , 2_j} + \ket{0_k , 2_j} \bra{2_k , 0_j} \big)  \\
         &\ \ \ +  \frac{ (3/2)^{1/2} (\ckku )^2}{\wku (\wo +2\wku )} \big( \ket{4_k , 0_j} \bra{0_k , 0_j} + \ket{0_k , 0_j} \bra{4_k , 0_j} \big) +
         \frac{ (3/2)^{1/2} (\cjjd )^2}{\wjd (\wo +2\wjd )} \big( \ket{0_k , 4_j} \bra{0_k , 0_j} + \ket{0_k , 0_j} \bra{0_k , 4_j} \big)\\
         &\ \ \ + \frac {\ckku \cjjd}{\wku +\wjd}\left( \frac 1{\wo +2\wku} + \frac 1{\wo +2\wjd} \right)\big( \ket{2_k , 2_j} \bra{0_k , 0_j} + \ket{0_k , 0_j } \bra{2_k , 2_j} \big) \Bigg\} .
    \end{split}
\end{equation}
\end{widetext}
(we remind our notation where, both in the ket and bra field states, the first element refers to subcavity 1 while the second one refers to subcavity 2).
All coefficients in (\ref{stato campi due modi}) containing a sum over field modes depend on the cutoff frequency. Importantly, the normalization factor $\Lambda^2$ defined in (\ref{Normalization factor}) is also strongly (quadratically) cutoff-dependent. We can evaluate its value for relevant values of the system parameters, typical of quantum optomechanics \cite{kharel2019high,kippenberg2007cavity} and circuit quantum electrodynamics (cQED) \cite{Butera25}:
for $\nu_0 = \wo /2\pi = 10^{10} \, \text{s}^{-1}$, $L_0 = 10^{-2} \, \text{m}$, $M = 10^{-8} \, \text{kg}$,
$\omega_P \simeq 10^{16} \, \text{rad} \cdot \text{s}^{-1}$, we find $\Lambda^2 \sim 10^{-24}$;
for $\nu_0 = 10^{10} \, \text{s}^{-1}$, $L_0 = 10^{-3} \, \text{m}$, $M = 10^{-30} \, \text{kg}$, $\omega_P = 10^{13} \, \text{rad} \cdot \text{s}^{-1}$, we find $\Lambda^2 \sim 10^{-8}$. The first case is typical of quantum optomechanics experiments and the second one is typical of cQED; they are also the same parameters $\nu_0 , \, L_0, \, M$ used in two of the plots (blue dots and red dots plots, respectively) of Fig. \ref{fig:neg_var_nmod_1D_num_new_code}. All these values show that $\Lambda^2$, that can be identified as our perturbative parameter, is indeed extremely small (similarly for the other plots in Fig. \ref{fig:neg_var_nmod_1D_num_new_code}), provided an appropriate cutoff frequency is introduced, giving a strong support to the validity of our perturbative expansion and its convergence.

In order to compute the negativity, we perform the partial transpose of $\rho_{kj}$ with respect to the mode $j$. At zeroth order, the partially transposed density matrix $\rho_{kj}^{t_j}$ is simply the projector onto the two-mode vacuum, $\rho_{kj}^{(0)} = \ket{0_k,0_j}\bra{0_k,0_j}$. All other terms constituting the partial transpose $\rho_{kj}^{t_j}$ are at least of second order in the coupling constants, that is $\rho_{kj}^{t_j} = \rho_{kj}^{(0)} + \delta R_{kj}$, where $\delta R_{kj}$ is $\mathcal{O}(C^2)$, with $C$ a collective symbol denoting the couplings $C_{kj}^{(\mu)}$. In other words, $\delta R_{kj}$ acts as a perturbation of the ``unperturbed'' partially transposed density operator $\rho_{kj}^{(0)}$. The unperturbed part has one nondegenerate eigenvalue corresponding to the vector $\ket{0_k} \otimes \ket{0_j}$, while every nonvacuum member of the Fock basis belongs to a degenerate subspace with zero unperturbed eigenvalue. At the lowest perturbative order, which in our case corresponds to $\mathcal{O}(C^2)$, the eigenvalues of $\rho_{kj}^{t_j}$ can therefore be obtained by restricting the perturbation to this
degenerate subspace, which is equivalent to considering $P_\perp^{(0)} \rho_{kj}^{t_j} P_\perp^{(0)}$, where $P_\perp^{(0)} = \mathbb{I} - \rho_{kj}^{(0)}$ is the projector onto
the subspace orthogonal to the two-mode vacuum state, in the spirit of first-order degenerate perturbation theory. At $\mathcal{O}(C^2)$, only the four basis states $\ket{1_k,0_j}$, $\ket{0_k,1_j}$, $\ket{2_k,0_j}$, and $\ket{0_k,2_j}$ carry nonvanishing contributing populations; the populations of $\ket{4_k,0_j}$, $\ket{0_k,4_j}$, and $\ket{2_k,2_j}$ appear at higher order and can be neglected at the order here considered. Therefore, in the ordered basis $\{\ket{1_k,0_j},\ket{0_k,1_j}, \ket{2_k,0_j},\ket{0_k,2_j}\}$, $P_\perp^{(0)} \rho_{kj}^{t_j} P_\perp^{(0)}$ can be expressed, without loss of generality, by a $4 \times 4$ matrix of the form
\begin{widetext}
\begin{equation}\label{block matrix}
\begin{pmatrix}
\bra{1_k,0_j}\rho_{kj}^{t_j}\ket{1_k,0_j} & 0 & 0 & 0 \\[6pt]
0 & \bra{0_k,1_j}\rho_{kj}^{t_j}\ket{0_k,1_j} & 0 & 0 \\[6pt]
0 & 0 & \bra{2_k,0_j}\rho_{kj}^{t_j}\ket{2_k,0_j}
& \bra{2_k,0_j}\rho_{kj}^{t_j}\ket{0_k,2_j} \\[6pt]
0 & 0 & \bra{0_k,2_j}\rho_{kj}^{t_j}\ket{2_k,0_j}
& \bra{0_k,2_j}\rho_{kj}^{t_j}\ket{0_k,2_j}
\end{pmatrix}.
\end{equation}
\end{widetext}
We observe that diagonal population terms remain positive after partial transposition and do not contribute to the negativity at this order. The only nontrivial block that can yield a negative eigenvalue is therefore the one spanned by the state vectors $\ket{2_k,0_j}$ and $\ket{0_k,2_j}$. This $2\times2$ block is generated by the partial transpose of the coherence $\ket{2_k,2_j}\bra{0_k,0_j}$, which becomes $\ket{2_k,0_j}\bra{0_k,2_j}$, and its Hermitian conjugate, which gives $\ket{0_k,2_j}\bra{2_k,0_j}$. Its entries involve only the diagonal couplings $C^{(1)}_{kk}$ and $C^{(2)}_{jj}$, with no sums over field modes: the negativity is therefore manifestly UV-finite, in contrast to the diagonal populations of $\ket{1_k,0_j}$ and $\ket{0_k,1_j}$, which do carry a cutoff dependence but do not enter into the negativity calculation at this order. Through a simple rotation we diagonalize the $2\times2$ matrix, obtaining two eigenvalues. One is always positive, while the other can be negative provided that
$(\wo +\wku +\wjd )/(\wku +\wjd )>1$; this condition is always verified since all frequencies involved are positive quantities. This clearly shows the presence of entanglement between the pair of field modes considered.

The complete expression of the negativity is
\begin{widetext}
    \begin{equation} \label{eq:asymmetric_neg}
    \begin{split}
        \mathcal{N}_{kj} = \frac{\hbar}{8 M l_1 l_2 \omega_0} \Bigg[& \sqrt{\Bigg( \frac{l_2}{l_{1}} \frac{\omega_{k}^2}{(\omega_0 + 2\omega_k)^2} - \frac{l_1}{l_{2}} \frac{\omega_{j}^2}{(\omega_0 + 2\omega_j)^2} \Bigg)^2 +  \frac{\omega_{k}^2 \omega_{j}^2}{(\omega_k + \omega_j)^2}  \Bigg( \frac{1}{\omega_0 + 2 \omega_k} + \frac{1}{\omega_0 + 2 \omega_j}  \Bigg)^2 } \\& \, \, \, - \frac{l_2}{l_{1}} \frac{\omega_{k}^2}{(\omega_0 + 2\omega_k)^2} - \frac{l_1}{l_{2}} \frac{\omega_{j}^2}{(\omega_0 + 2\omega_j)^2}  \, \Bigg] \, ,
    \end{split}
\end{equation}
\end{widetext}
where from now onwards, we simplify the notation by setting $\wk = \wku$ and $\wj = \wjd$; $l_1$ and $l_2$, as defined at the beginning of Sec. \ref{sec:model}, are respectively the distances of the equilibrium position of the movable wall from the left and right fixed walls.
We wish to emphasize that Eq. (\ref{eq:asymmetric_neg}), one of our main results, is finite and cutoff independent.
We also note that the expression of the negativity obtained above coincides with the one that would be obtained if, instead of considering the reduced state of the field on two modes (one for each cavity), a single-mode approximation in each cavity would have been used, discarding since the beginning all other field modes in the subcavities.

\begin{figure}[h!]
    \centering
    \includegraphics[width=0.48\textwidth]{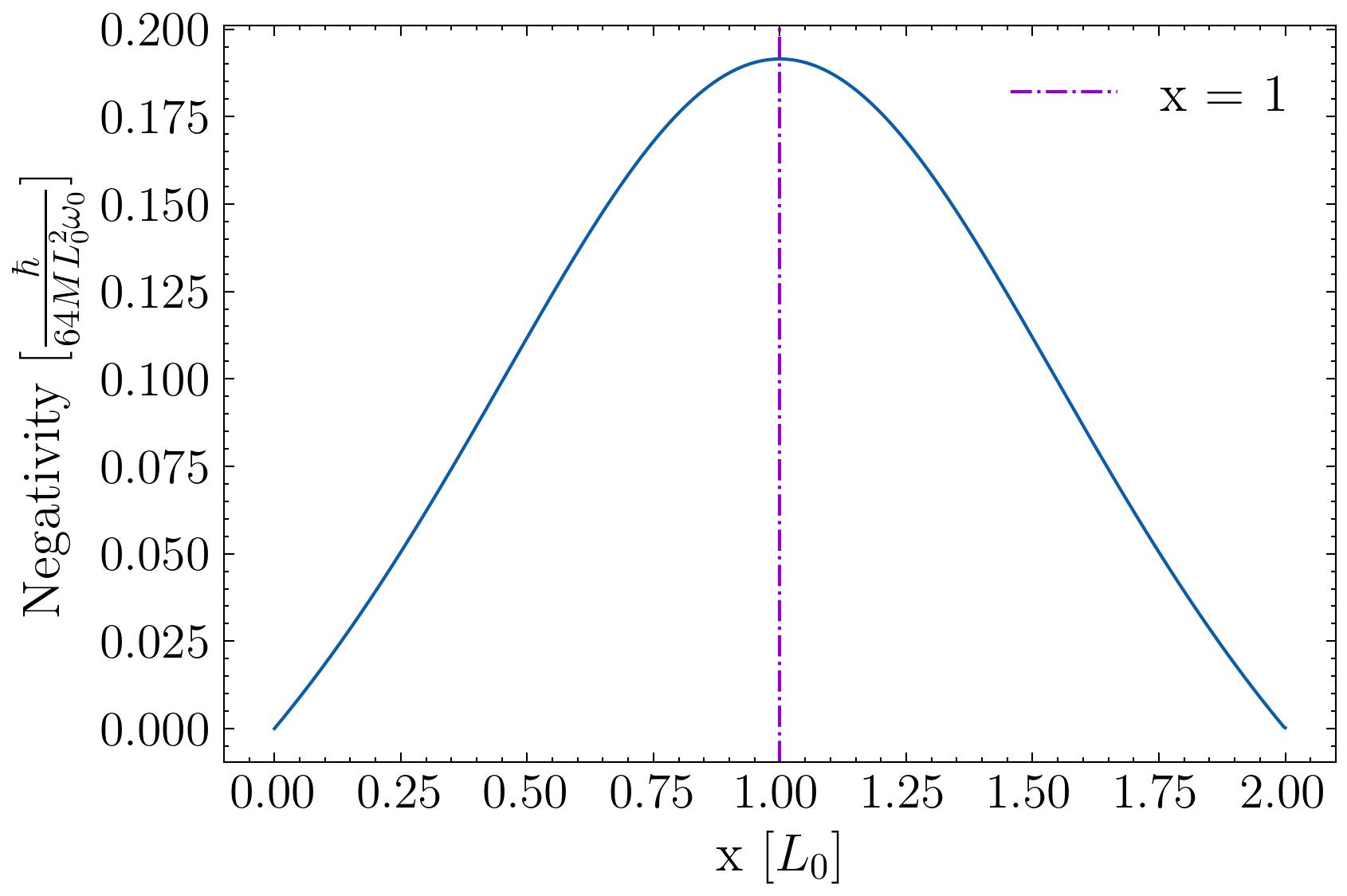}
    \caption{Negativity (\ref{eq:asymmetric_neg}), expressed in units of the dimensionless factor $\hbar/(64ML_0^2 \wo)$, as a function of $x$, the wall's equilibrium position in units of $L_0 \,$ $(0 < x < 2)$.}
    \label{fig:neg_sm_approx_1D_ana_var_eq_pos}
\end{figure}

We now discuss the physical properties of the entanglement between field modes in our system that can be obtained from our result (\ref{eq:asymmetric_neg}) for the negativity, with particular emphasis on their dependence from  relevant parameters of the system.

\begin{figure}[h!]
    \centering
    \includegraphics[width=0.48\textwidth]{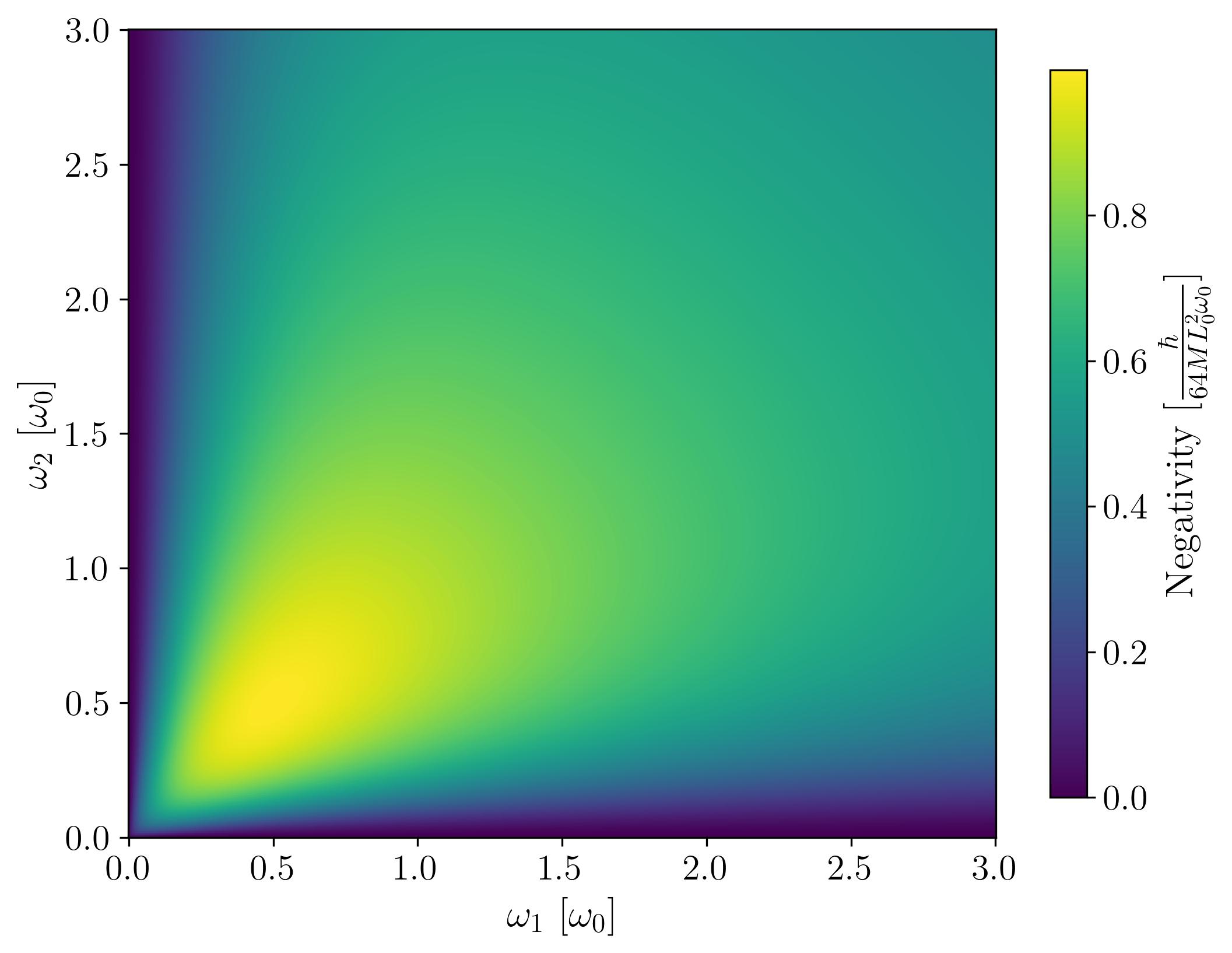}
    \caption{Negativity (\ref{eq:asymmetric_neg}) as a function of
    $\omega_1 \equiv \omega_k = (\pi ck)/l_1$ and $\omega_2 \equiv \omega_j = (\pi cj)/l_2$, in units of $\omega_0$, for the symmetric configuration $l_1 = l_2 = L_0$. The negativity is expressed in units of the dimensionless factor $\hbar/(64ML_0^2 \wo)$.}
    \label{fig:neg_sm_approx_3D}
\end{figure}

\begin{figure}[h!]
    \centering
    \includegraphics[width=0.48\textwidth]{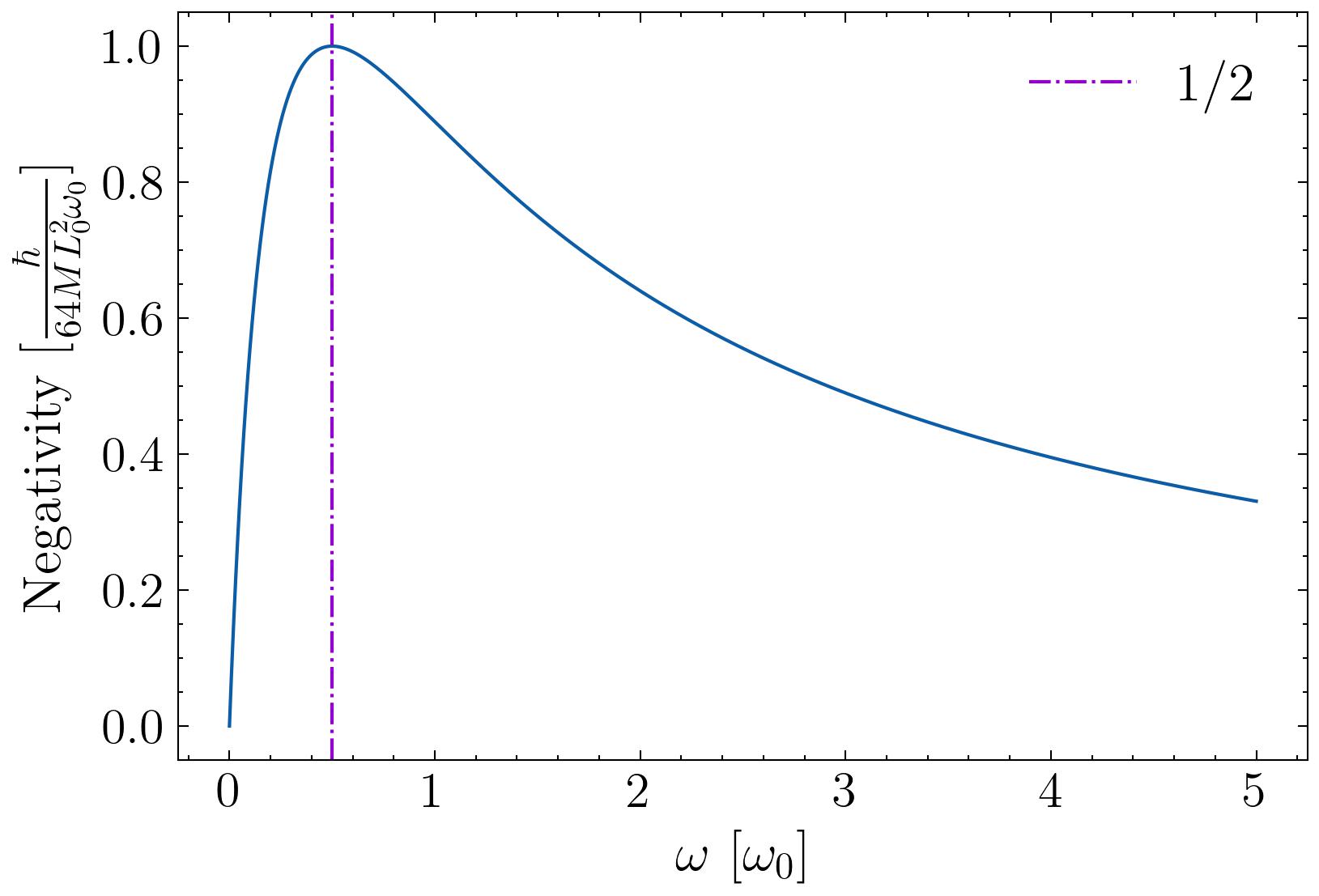}
    \caption{Negativity in Eq. (\ref{eq:symmetric_neg}) as a function of
    $\omega = (\pi ck)/L_0$, in units of $\omega_0$, for the symmetric configuration $l_1 = l_2 = L_0$. The negativity is expressed in units of the dimensionless factor $\hbar/(64ML_0^2 \wo)$.}
    \label{fig:neg_sm_approx_1D_ana}
\end{figure}

In Fig. \ref{fig:neg_sm_approx_1D_ana_var_eq_pos} we plot the negativity, in units of the dimensionless scale factor $\hbar / (64 M \omega_0 L_0^2)$,
as a function of the wall's equilibrium position $l_1$ normalized to half the distance $L_0$ between the two fixed walls, namely $x=l_1/L_0 \in (0,2)$; $x=1$ corresponds to the equilibrium position of the movable wall at the midpoint between the fixed walls. The figure clearly shows that the negativity is maximized for a symmetric geometric configuration, $x=1$, that is $l_1=l_2=L_0$. For this reason we now concentrate on this specific symmetric configuration of the three walls.
We wish to mention that in Figures \ref{fig:neg_sm_approx_1D_ana_var_eq_pos}, \ref{fig:neg_sm_approx_3D}, \ref{fig:neg_sm_approx_1D_ana}, for ease of visualization our discrete mode frequencies are idealized as continuous variables.
In Fig. \ref{fig:neg_sm_approx_3D} the negativity (\ref{eq:asymmetric_neg}) is plotted as a function of $\omega_1 =\omega_k / \omega_0$ and $\omega_2 =\omega_j / \omega_0$, and in Fig. \ref{fig:neg_sm_approx_1D_ana} as a function of $\omega / \omega_0$ in the case of equal-frequency modes in the two subcavities, $\omega_k = \omega_j \equiv \omega$.
It is evident from these plots that, although the entanglement between the fields in the two subcavities is nonzero everywhere, the negativity between pairs of field modes is maximized (and thus the entanglement is the strongest) whenever $\omega_k / \omega_0 \lesssim 1$ and $\omega_j / \omega_0 \lesssim 1$.
Also, the maximum is obtained for $\omega_k = \omega_j = \omega_0 / 2$.
This result is expected. In fact, under the condition $\omega_k = \omega_j = \wo /2$, analogous to a resonance, the exchange of virtual excitations between the wall and the field is most effective, leading to a stronger accumulation of intercavity correlations and consequently to the maximum value of the negativity (see also Eq. (\ref{eq:symmetric_neg}) below) \cite{Ayoub-Akram21}.
Also, this condition strongly resembles that occurring in the dynamical Casimir effect for the emission of pairs of real photons, although in our case we deal with virtual quanta; thus it is a sort of weak analog to the resonance condition of real photon pairs generated in the dynamical Casimir effect \cite{Butera-Passante13}. In this last case the negativity (\ref{eq:asymmetric_neg}) reduces to
\begin{equation} \label{eq:symmetric_neg}
    \mathcal{N}(\omega) = \frac{\hbar \omega}{8 M L_{0}^2 (\omega_0 + 2 \omega)^2}
    = \frac{\hbar \omega /\wo}{8 M L_{0}^2 \wo (1 + 2 \omega /\wo)^2}
    \, .
\end{equation}

Eqs. (\ref{eq:asymmetric_neg}) and (\ref{eq:symmetric_neg}) clearly show that the value of the negativity depends on the parameters $\omega_0$, $L_0$ and $M$ through the dimensionless factor
$\hbar M^{-1} \wo^{-1}L_0^{-2}$, which is directly related to the quantum position fluctuations of the finite-mass movable wall (of the order of $\Delta x^2 \sim \hbar M^{-1} \wo^{-1}$); this highlights the genuinely quantum origin of the entanglement between the field modes, mediated by the quantum fluctuations of the movable wall, that we find. It is thus of interest to investigate the dependence of the negativity from this triad of parameters, in particular using values reachable in current or foreseeable optomechanics experimental setups.

Using for example $\nu_0 =\omega_0/2\pi = 10 \hspace{0.1cm} \mathrm{GHz}, \hspace{0.1cm} L_0 = \hspace{0.1cm} 10^{-2 } \, \mathrm{m}, \hspace{0.1cm} M = 10^{-8} \hspace{0.1cm} \mathrm{kg}$, that are numerical values of the order of those obtained in \cite{kharel2019high} exploiting bulk acoustic phonons in a quartz crystal inside a macroscopic high-frequency cavity, and setting $\omega_k = \omega_j \sim \wo /2$ (this condition, as already shown, maximizes the negativity), from (\ref{eq:symmetric_neg}) we obtain  an absolute value of the negativity of the order of $10^{-35}$, that is a quite small value.
A possibility to increase the absolute value of the negativity is to reach a smaller mass $M$ of the movable wall or to decrease $L_0$, or both.
Quite smaller values of mass and cavity length have been indeed obtained in optomechanics experiments, see for example \cite{Aspelmeyer-Kippemberg14,villafane2018quantum}. Also, in \cite{Liu:13} a mass below the picogram scale has been obtained. For a review of optomechanics systems with levitated micro and nanoparticles with very low mass and dissipation, see \cite{Millen_2020}.
In our case, taking reasonable values such as $M \sim 10^{-18}$ kg, $L_0 \sim 10^{-6}$ m and
$\nu_0 = \wo /2\pi \sim 10^2$ GHz,
the dimensionless factor $\hbar / (M \omega_0 L_0^2)$ becomes quite large yielding a much larger negativity, of the order of $10^{-21}$. Hence, a factor of the order of $10^{14}$ is gained with respect to the previous case, even if such values give a minimum mode frequency $\wk = \wj =\pi c/L_0 \sim 10^{15} \, \text{s}^{-1} \gg \wo$, and thus do not satisfy at all the ``resonance'' condition $\wk = \wj = \wo /2$ that maximizes the negativity for given values of $(M,\, \wo ,\, L_0)$. Furthermore, mass $M$ and frequency $\omega_0$ can be reduced both, or at least one can be lowered without much affecting the other, by decoupling the effective mass from the mechanical oscillator's stiffness. This can be achieved by reducing lateral dimensions \cite{zheng2012femtogram}, or reducing mechanical dissipation \cite{tsaturyan2017ultracoherent}, or confining mechanical modes to ultra-small volumes \cite{kirchhof2021tunable} (lowering $M$ and enabling tuning of $\omega_0$), or combining optical and mechanical confinement to maximize optomechanical coupling while controlling $M$ and $\omega_0$ \cite{li2015optomechanical,huang2015strong,leijssen2015strong}. Finally, we wish to mention that circuit-QED devices as those considered in \cite{Butera-Carusotto19} seem a promising setup for further increasing the negativity by several orders of magnitude; we hope to address this and related aspects in a future publication.

\begin{figure}[h!]
    \centering
    \includegraphics[width=0.48\textwidth]{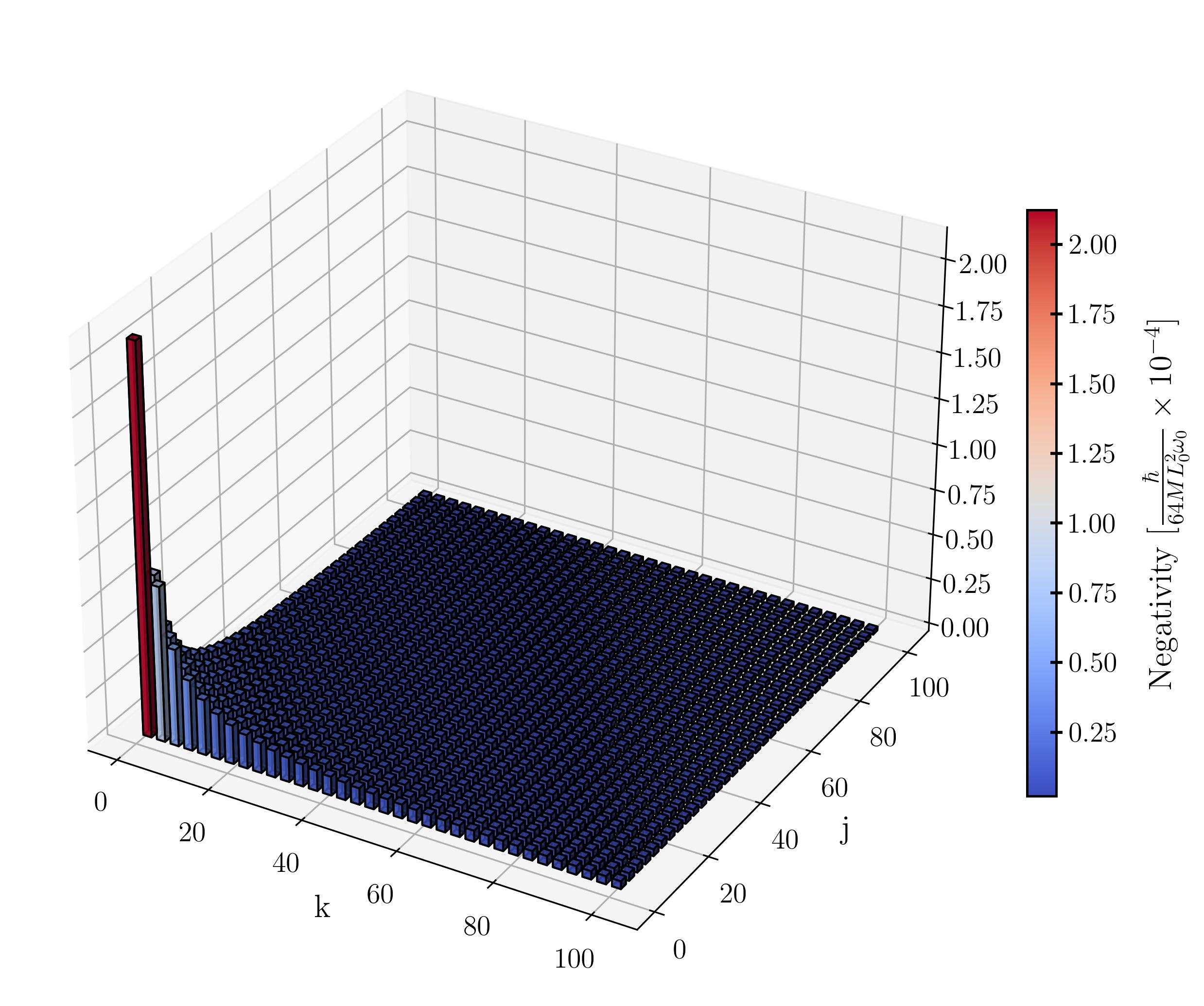}
    \caption{3D histogram of the negativity in Eq. (\ref{eq:asymmetric_neg}), quantifying the entanglement between pairs of modes $k$ and $j$, in terms of the dimensionless parameter $\hbar /(64ML_0^2\wo)$.}
    \label{fig:neg_entanglement_bw_modes}
\end{figure}

Another way to visualize which field mode pairs are more entangled is by depicting the negativity associated to them as a three-dimensional histogram as a function of $k, j \in \mathbb{N}^+$, as shown in Fig. \ref{fig:neg_entanglement_bw_modes}, where the numerical values for the parameters in the triad are of the order of those attained in \cite{villafane2018quantum}. This parametric plot shows that the field modes that become most entangled across the two subcavities are the ones for which $k = j = 1$.

\section{Numerical analysis of the multimode negativity}
\label{sec:entanglement2}

In the previous section we quantified the entanglement in our two-cavity system by evaluating the negativity between two single field modes when the system is in its interacting ground state, and we found that they are entangled; however, the negativity, for realistic values of all parameters involved, is quite small.

In this section we extend our evaluation of the negativity to the case of a larger number of modes in the two subcavities. We
numerically evaluate the entanglement (negativity) between the fields in the two subcavities, each containing a finite number of modes $N_{\mathrm{mod}}$. The starting point is the field density operator (\ref{reduced density operator}). In this case, although from Eq. (\ref{eq:symmetric_neg}) the contribution of a given mode decreases after we go beyond the modes for which $\omega \approx \omega_0/2$, many modes can still be taken into account.
The number of modes depends on both $L_0$, that determines their spacing, and the upper frequency limit given by the cutoff, that is the plasma frequency $\omega_P$ for a metal movable mirror;
this number is given by Eq. (\ref{number modes}).
The dimension of the system's Hilbert space scales as $N_{\mathrm{exc}}^{2 N_{\mathrm{mod}}}$, where $N_{\mathrm{exc}}$ is the finite number of excitations in each field mode.
In the present second-order perturbative treatment, $N_{\mathrm{exc}} = 5$ (the ground state and the states with one or two pairs of excitations in the two subcavities). A direct diagonalization of the partially transposed density matrix would become intractable as $N_{\mathrm{mod}}$ increases.
The structure discussed around Eq. (\ref{block matrix}) provides the starting point for identifying the sectors that, at the order considered, can contribute negative eigenvalues in the multimode case.
In the general case of $N_{\text{mod}}$ field modes per subcavity, at the perturbative order considered only two blocks
of the partially transposed density matrix can contain negative eigenvalues and are therefore relevant to the negativity calculation.
Each of these blocks is indexed by two pairs of mode indices, $(j,k)$ and $(l,m)$. Since each pair can take $N_{\text{mod}}^2$ values, each block is a $N_{\text{mod}}^2 \times N_{\text{mod}}^2$ matrix and
therefore contains $N_{\text{mod}}^4$ elements. The numerical calculation can thus be restricted to the diagonalization of these two blocks, corresponding to a total of $2N_{\text{mod}}^4$ relevant matrix elements,
rather than to the diagonalization of the full partially transposed density matrix. This is the multimode generalization of the restriction to the nontrivial sector discussed for a single pair of modes around Eq. (\ref{block matrix}), and it drastically reduces the numerical problem with respect to the full density matrix, which would contain $5^{4N_{\text{mod}}}$ elements within the excitation-number truncation considered here.
For a typical plasma frequency $\omega_P \sim 10^{16} \, \text{rad} \text{s}^{-1}$ and a small cavity with a length of the order of micrometers, we can estimate $N_{mod} \sim 10$, while a larger cavity with $L_0$ in the centimeter range would roughly contain $10^5$ relevant modes.
We wish to stress, however, that considering a typical case of cQED, which is a very promising candidate for implementing our two-cavity system,
$\omega_{cutoff} \sim 10^{12} - 10^{13} \, \text{rad} \cdot \text{s}^{-1}$, or even less, yielding a much smaller number of modes, in the range of $10-100$.

\begin{figure}[h!]
    \centering
    \includegraphics[width=0.48\textwidth]{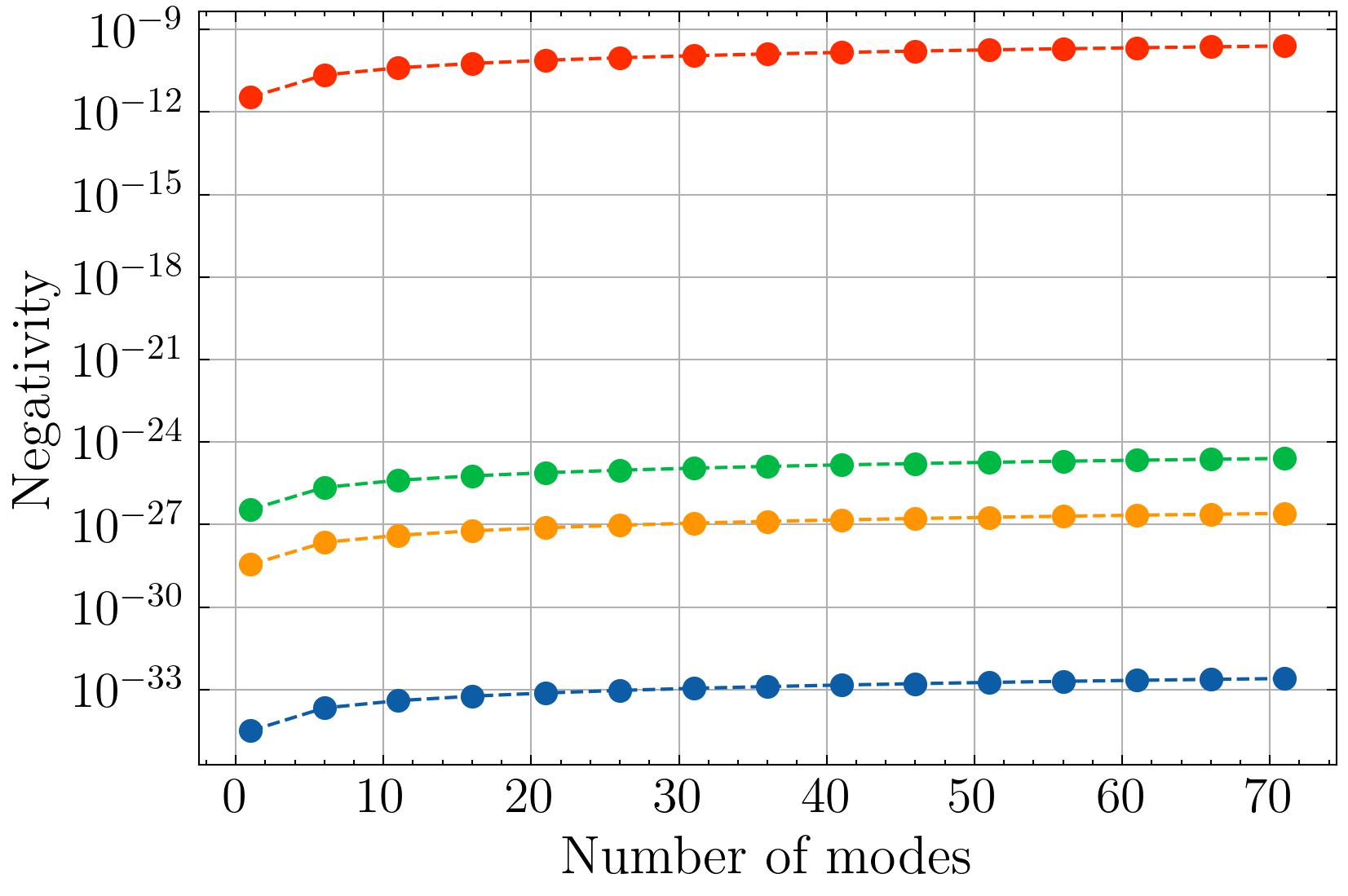}
\caption{Semilog plot of the negativity calculated numerically for different parameters: Blue dots refer to:
$\nu_0 = 10^{10} \, \text{s}^{-1}, \hspace{0.1cm} L_0 = 10^{-2} \, \text{m}, \hspace{0.1cm} M=10^{-8} \, \text{kg}$;
Green ones: $\nu_0 = 10^{11} \, \text{s}^{-1}, \hspace{0.1cm} L_0 = 10^{-6} \, \text{m}, \hspace{0.1cm} M=10^{-12} \, \text{kg}$;
Yellow dots: $\nu_0 = 10^{7} \, \text{s}^{-1}, \hspace{0.1cm} L_0 = 10^{-5} \, \text{m}, \hspace{0.1cm} M=10^{-11} \, \text{kg}$;
Red dots: $\nu_0 = 10^{10} \, \text{s}^{-1}, \hspace{0.1cm} L_0 = 10^{-3} \, \text{m}, \hspace{0.1cm} M=10^{-30} \, \text{kg}$.}
    \label{fig:neg_var_nmod_1D_num_new_code}
\end{figure}

The quantitative analysis of this multimode case has been performed numerically due to the complexity of the reduced density operator (\ref{reduced density operator}). We have performed numerically the analysis only for the first $\sim 70$ modes of the fields, due to the high computational demand of the algorithm needed for the computation. This is quite a good number of modes for the case of a micrometer-sized cavity as that considered in \cite{villafane2018quantum} and \cite{kippenberg2007cavity}, even if it is far from the $10^5$ relevant modes required for describing the case of a centimeter-range cavity as the optomechanical cavity used in \cite{kharel2019high}.
On the other hand, this number of modes is fully adequate for a typical cQED case, where a number of modes in the range of $10-100$ is expected, as discussed above.
The results of our numerical analysis for the negativity are shown in Fig. \ref{fig:neg_var_nmod_1D_num_new_code} for different realistic values of the parameters $\omega_0, \hspace{0.1cm} L_0, \hspace{0.1cm} M$ as a function of $N_{\mathrm{mod}}$.
In particular:
the blue-dots curve is for  $\nu_0 = 10^{10} \, \text{s}^{-1}, \hspace{0.1cm} L_0 = 10^{-2} \, \text{m}, \hspace{0.1cm} M=10^{-8} \, \text{kg}$ (numerical values of the order of magnitude of those used in \cite{kharel2019high});
the green-dots curve is for  $\nu_0 = 10^{11} \, \text{s}^{-1}, \hspace{0.1cm} L_0 = 10^{-6} \, \text{m}, \hspace{0.1cm} M=10^{-12} \, \text{kg}$ (numerical values of the order of magnitude of those used in \cite{villafane2018quantum});
the yellow-dots curve is for  $\nu_0 = 10^{7} \, \text{s}^{-1}, \hspace{0.1cm} L_0 = 10^{-5} \, \text{m}, \hspace{0.1cm} M=10^{-11} \, \text{kg}$ (numerical values of the order of magnitude of those used in \cite{kippenberg2007cavity});
the red-dots curve is for  $\nu_0 = 10^{10} \, \text{s}^{-1}, \hspace{0.1cm} L_0 = 10^{-3} \, \text{m}, \hspace{0.1cm} M=10^{-30} \, \text{kg}$ (numerical values of the order of magnitude of possible values mentioned for the circuit QED analogous system considered in \cite{Butera-Carusotto19}).

In the cQED case, the relevant cutoff frequency is the plasma frequency of the SQUID device simulating the movable wall, which is of the order of $\omega_p \sim 10^{11}-10^{13} \, \text{rad} \cdot \text{s}^{-1}$, far less than a typical plasma frequency of a metal. Hence, far fewer modes are available, and, as already shown in Sec. \ref{sec:entanglement}, the parameter $\Lambda^2$ is far less than one also in this case. Moreover, in this case the number of modes is typically less than $100$, as already mentioned.

We observe that the plots in Fig. \ref{fig:neg_var_nmod_1D_num_new_code}
qualitatively indicate that the main contribution comes from roughly the first $\sim 30$ modes and that the inclusion of the other modes considered yields an increase in the negativity of about two orders of magnitude, showing also a plateau when the number of modes is increased. The remaining modes, when present, seem to lead only to a small increase in the negativity of less than one order of magnitude, which suggests the approach of an asymptote when increasing the total number of field modes. Even if this is only a qualitative indication, inferred on the basis of the (small) number of modes considered, it becomes of particular relevance and value for the red-dots plot, which has parameter values typical of cQED and thus contains a much lower number of modes (less than about $100$, as mentioned before).

\section{Conclusions}
\label{sec:conclusions}

We have considered a one-dimensional two-cavity system consisting of a movable conducting wall placed between two fixed conducting plane boundaries. Two massless scalar fields are defined, one in each cavity. The mechanical degrees of freedom of the movable wall, assumed to have a very low mass and bound to its equilibrium position by a harmonic potential, have been treated quantum-mechanically. As it is known, the motion of the movable wall induces an effective interaction between the field modes and an effective fields-wall interaction, described by the Law Hamiltonian \cite{Law94,Law95}, as well as correlations between field observables in the two cavities \cite{Montalbano-Armata23}. We have calculated the interacting ground state of the system up to the second order in the interaction between the two scalar fields and the movable boundary, and the reduced field density operator.
An appropriate upper cutoff frequency, motivated by the fact that the metallic mirrors become transparent above their plasma frequency, has been introduced to eliminate ultraviolet divergences and obtain convergence of the perturbative approach.
We have analyzed the entanglement between field modes in the two cavities by evaluating the negativity using both analytical and numerical approaches, both in the case of two single modes, one in each of the two subcavities, and between a larger number of field modes in each of the two subcavities.
Our results show that the quantum position fluctuations of the movable wall generate entanglement between the field modes belonging to opposite subcavities. We have also discussed the behavior of the negativity as a function of all relevant parameters of the system, such as the frequency of the field modes considered, the size of the two subcavities, the mass and oscillation frequency of the movable wall, as well as the number of field modes considered.

\begin{acknowledgments}
The authors acknowledge financial support from the FFR2023 (R.P. and L.R.) and FFR2024 (L.R.) grants from the University of Palermo, Italy. L.G.C. acknowledges support from PR FSE+ Sicilia 2021-2027.  PNRR MUR Project MINTQT Partenariato Esteso NQSTI PE00000023  Spoke 9 - CUP: E63C22002180006 is also acknowledged.
\end{acknowledgments}

\section*{DATA AVAILABILITY}

There are no publicly available research data or software supporting this manuscript. Requests for further information or data should be sent to the authors.

\bibliography{biblio.bib}

\end{document}